\documentclass[10pt,conference]{IEEEtran}
\IEEEoverridecommandlockouts
\usepackage{cite}
\usepackage{amsmath,amssymb,amsfonts}
\usepackage{algorithmic}
\usepackage{graphicx}
\usepackage{textcomp}
\usepackage{url}
\usepackage{multirow}
\usepackage{tcolorbox}
\usepackage{booktabs}
\usepackage{tcolorbox}
\usepackage{multirow}
\usepackage{bbding}
\usepackage{utfsym}
\usepackage{ulem}
\usepackage{balance}
\usepackage{wasysym}
\usepackage{hyperref}
\usepackage{multirow}
\usepackage{makecell}
\usepackage{pifont}

\def\eg{\textit{e.g.,} }
\def\ie{\textit{i.e.,} }

\newtcolorbox{AcademicBox}[1][]{academicbox=#1}
\definecolor{SoftBlue}{RGB}{135, 206, 250}  
\definecolor{SoftOrange}{RGB}{255, 224, 178} 
\definecolor{SoftGreen}{RGB}{144, 238, 144}  
\definecolor{CorrectGreen}{RGB}{76, 175, 80} 
\definecolor{ErrorRed}{RGB}{211, 47, 47} 

\def\BibTeX{{\rm B\kern-.05em{\sc i\kern-.025em b}\kern-.08em
    T\kern-.1667em\lower.7ex\hbox{E}\kern-.125emX}}

\begin{document}

\title{An Evaluation of Requirements Modeling for Cyber-Physical Systems via LLMs}


\author{\IEEEauthorblockN{Anonymous Author(s)}}

 \author{
 \IEEEauthorblockN{Dongming Jin$^{1,2}$, Shengxin Zhao$^{4}$, Zhi Jin*$^{1,2}$, Xiaohong Chen$^{3}$, Chunhui Wang$^{4}$, Zheng Fang$^{1,2}$, Hongbin Xiao$^{5}$}
    \IEEEauthorblockA{$^1$ School of Computer Science, Peking University, China}
    \IEEEauthorblockA{$^2$ Key Lab of High-Confidence of Software Technologies (PKU), Ministry of Education, China}
    \IEEEauthorblockA{$^3$ Shanghai Key Laboratory of Trustworthy Computing, East China Normal University, China}
    \IEEEauthorblockA{$^4$ College of Computer Science and Technology, Inner Mongolia Normal University, China}
    \IEEEauthorblockA{\texttt{dmjin@stu.pku.edu.cn}, \texttt{zhijin@pku.edu.cn}}
}


\maketitle

\begin{abstract}
Cyber-physical systems (CPSs) integrate cyber (\eg computation and communication) and physical components (\eg sensors and actuators) and enable them to interact with each other to meet user needs. The needs for CPSs span rich application domains such as healthcare and medicine, smart home, smart building, etc. This indicates that CPSs are all about solving real-world problems. With the increasing abundance of sensing devices and effectors, the problems wanted to solve with CPSs are becoming more and more complex. It is also becoming increasingly difficult to extract and express CPS requirements accurately. Problem frame approach aims to shape real-world problems by capturing the characteristics and interconnections of components, where the problem diagram is central to expressing the requirements. CPSs requirements are generally presented in domain-specific documents that are normally expressed in natural language.
There is currently no effective way to extract problem diagrams from natural language documents. CPSs requirements extraction and modeling are generally done manually, which is time-consuming, labor-intensive, and error-prone.

Large language models (LLMs) have shown excellent performance in natural language understanding. It can be interesting to explore the abilities of LLMs to understand domain-specific documents and identify modeling elements, which this paper is working on. To achieve this goal, we first formulate two tasks (\ie entity recognition and interaction extraction) and propose a benchmark called \textit{CPSBench}. Based on this benchmark, extensive experiments are conducted to evaluate the abilities and limitations of seven advanced LLMs. We find that (1) LLMs have limited ability to model the requirements for CPSs using problem diagrams. (2) LLMs have a better understanding of general concepts than specialized concepts. (3) The performance of LLMs can be improved in a few-shot setting. Finally, we establish a taxonomy of LLMs hallucinations in CPSs requirements modeling using problem diagrams. 
These results will inspire research on the use of LLMs for automated CPSs requirements modeling.
\end{abstract}

\section{Introduction}
Cyber-physical systems (CPSs) are pervasive in modern life~\cite{nguyen2017model}, from mobile phones and other electronic products to cars and spacecraft~\cite{menghi2020approximation}. CPSs are characterized by the tight coupling of physical environments and software components~\cite{xu2023semantics} to allow the software to interact with the physical environments~\cite{mandrioli2023stress}. 
With the emergence of various sensing and actuating devices, CPSs continue to grow in size and complexity~\cite{frepa} and the interactions between software components and their physical environments are becoming more complex~\cite{stadler2022flexible}. 
This poses a significant challenge in obtaining accurate requirements and ensuring that the CPSs meet the expected functionality and performance~\cite{BouskelaFGJOTT22}. 

Problem frame (PF) approach~\cite{jackson2005problem} aims to shape real-world problems by capturing the characteristics and interconnections of components. It treats the interactive environments as a first-class citizen, emphasizing the interactions between software and its environments, and is therefore particularly suitable for modeling the CPSs requirements~\cite{JinRE2019, jin_environment_2018}. However, CPSs requirements modeling in terms of PF (mostly problem diagram) requires tremendous human efforts. 
The efforts rely on the understanding of the domain documents in natural language to identify the required modeling elements scattered throughout the documents, which is time-consuming, labor-intensive, and error-prone.
If this process can be automated, it will greatly improve the efficiency of requirements modeling.

\begin{figure}
  \centering
  \includegraphics[width=\linewidth]{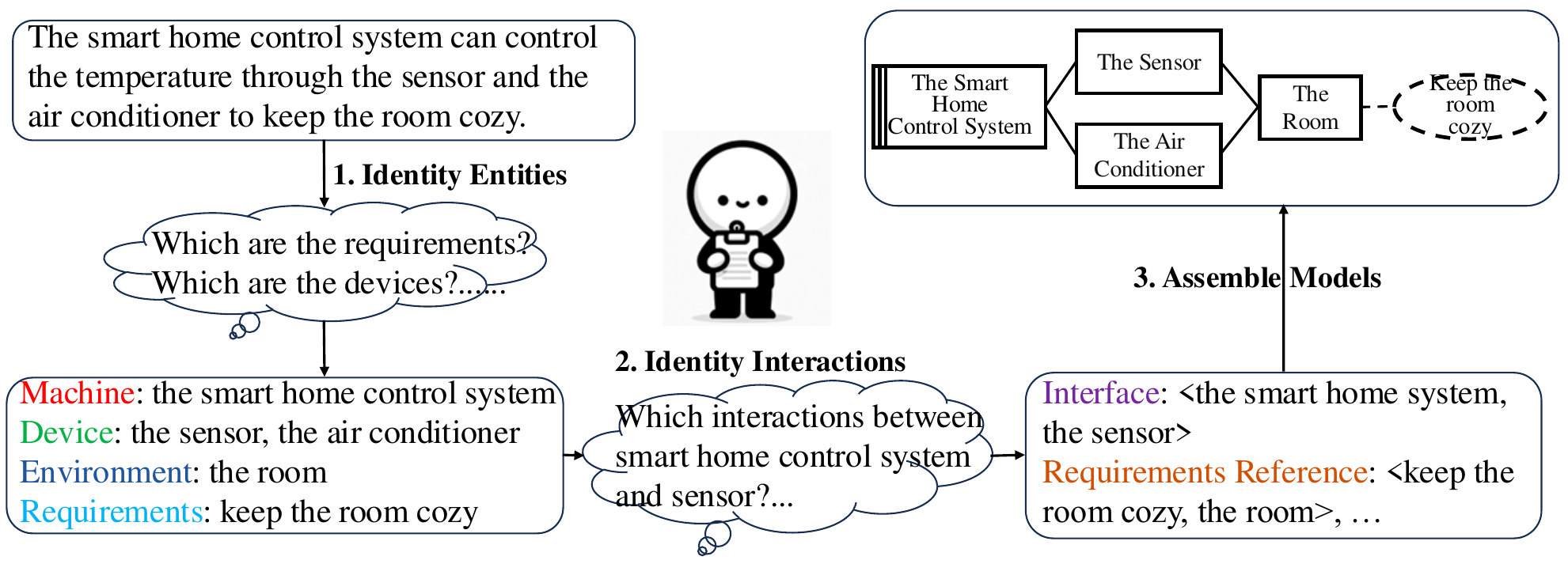}
  \caption{The process of requirements modeling for CPSs from natural language requirement document by developers}
  \label{fig:teaser}
\end{figure}

There have been some works to try to automatically understand natural language requirements to assist in requirements models generation~\cite{rajbhoj2023doctomodel}~\cite{zaki2022rcm}~\cite{javed2021imer}.  
Recently, large language models (LLMs) such as ChatGPT~\cite{gpt-3.5} have demonstrated excellent performance in natural language understanding. Some recent works have explored LLMs ability to understand requirements and accomplish various Requirements Engineering (RE) tasks~\cite{arora2024advancing}, such as requirements completeness~\cite{luitel2024improving}, specification generation~\cite{lutze2024generating}, and inconsistency detection~\cite{fantechi2023inconsistency}. 
Investigations on requirements modeling using LLMs are also involved~\cite{camara2023assessment}.
For example, recent works~\cite{ferrari2024model}~\cite{chen2023automated}~\cite{RuanCJ23} are trying to use LLMs to help construct the sequence diagrams, the goal models, and the problem diagrams respectively. Modeling the sequence diagram focuses on eliciting the behavior concern, while the goal modeling emphasizes understanding the hierarchical structure of requirements. 
Unlike them, which focus on a single concern, constructing the problem diagrams requires understanding multiple concerns about both the environment and the interactions (behavior). This should be more difficult. A systematic evaluation of the capabilities of LLMs on understanding domain requirements documents, extracting PD modeling elements, and constructing requirements models for CPSs is very necessary to explore. 

This paper proposes to conduct an evaluation to investigate the performance and limitations of various advanced LLMs on CPSs requirements modeling. To achieve this goal, we first construct a benchmark based on the real-world requirements documents. This is not trivial due to the following challenges: \textbf{(1) Difficulty in accessing requirements documents in the real world.} These documents are often private for enterprises and tend not to be public. Existing works~\cite{al2018use}~\cite{nguyen2015rule} are generally done by using the cases from books or courses for evaluation. There is a gap in scale and complexity between these cases and real-world requirements documents. For example, a real-world embedded system requirements document can be dozens of pages long~\cite{frepa}, but the evaluation cases for IT4RE~\cite{al2018use} do not exceed ten sentences. It is necessary to collect requirements documents to increase the size and complexity of the evaluation, reducing the gap with real situations. \textbf{(2) Expensive human effort.} Sufficient prior knowledge is required to complete the requirements modeling task. It is necessary to leverage manual annotation to guide the construction of the benchmark. However, the manual annotation process requires experienced analysts to spend a significant amount of time understanding the requirements documents thoroughly. It is very expensive to annotate requirements documents. (3) \textbf{Noisy data}. Existing requirements documents often contain noisy information such as incomplete sentences and unreadable tables. This noisy information makes it difficult to process documents.

In this paper, we formalize the task of requirement modeling (\eg problem diagram construction) into two types of identification tasks, including entity recognition and interaction extraction. It is inspired by the process of manually building requirements models from documents, as shown in Figure \ref{fig:teaser}. Specifically, given a requirement description, analysts first recognize entity elements and determine their types (\eg \textit{machine} in problem diagrams). Then, they determine whether there are interactions among the entity elements (\eg \textit{interface} in problem diagrams) and finally decide the interactions to make and the constraints among the interactions. 

We collect multiple CPSs requirements documents in natural language and propose a \textbf{CPS}s requirements modeling \textbf{Bench}mark called {\sc CPSBench}. The {\sc CPSBench} consists of 12 enterprise-level requirements documents and 30 tutorial cases. Each sample in {\sc CPSBench} includes four parts: requirements, Entity, Interaction, and Problem Diagram. 

We apply the few-shot reasoning strategy~\cite{xie2021explanation} to evaluate the capabilities and limitations of seven advanced LLMs (in Section \ref{sec:llms}) on CPS requirements modeling with problem diagrams using our {\sc CPSBench}. Our evaluation finds that (1) LLMs do not achieve sufficient effectiveness in modeling the CPSs requirements using problem diagrams for practical applications. Current LLMs achieve a recall rate of only around 60\%, failing to recognize almost half of the modeling elements. (2) LLMs have a better understanding of general requirements concepts than specialized concepts. Specifically, LLMs have a richer knowledge of the machine domain (E-MD), physical devices (E-PD), environmental entity (E-EE), and interface (R-IN). However, LLMs lack knowledge of concepts such as design domain (E-DD), requirements domain (E-RD), requirements reference (R-RR), and requirements constraints (R-RC). The meanings of these concepts can be found in our task definition (Section \ref{sec:definition}) (3) LLMs can improve their performance with more shots in the prompt.

Finally, we conduct a comprehensive analysis of the modeling elements recognized by LLMs from the natural language requirements documents and establish a taxonomy of LLMs hallucinations in CPSs requirements modeling. Additionally, we discuss the directions for improving CPSs requirements modeling via LLMs in the future.

We summarize our contributions as follows.
\begin{itemize}
    \item We propose a CPSs requirements modeling benchmark named {\sc CPSBench}. The {\sc CPSBench} consists of requirements documents in the real world, reducing the gap between evaluation and practical application.
    \item We conduct an extensive evaluation of CPSs requirements modeling for seven popular LLMs and gain some insights into their strengths and limitations. 
    \item We establish a taxonomy of LLMs hallucinations in requirements modeling for CPSs and provide directions for improvement in the future.
\end{itemize}

\textbf{Data Availability}. We open-source our replication package~\cite{web:code}, which includes the benchmark {\sc CPSBench} and the source code of evaluation, to enable other researchers and practitioners to replicate our work and validate their studies.

In the remainder of the paper, section \ref{sec:definition} illustrates the formalization of the CPS requirements modeling. Section \ref{sec:benchmark} introduces the process of constructing the benchmark. Section \ref{sec:approach} presents the reasoning approach for LLMs. Section \ref{sec:studydesign} sets up the experiments. Section \ref{sec:resultandanalysis} describes the results and analysis. Section \ref{sec:discussion} provides the discussion. Section \ref{sec:relatedworks} reviews the related works. Section \ref{sec:conclusion} concludes this paper.

\section{Task Definition} \label{sec:definition}
In this section, we illustrate the formulation of the CPS requirements modeling. We define the overview of our formulation and describe the tasks in the subsequent sections.

\subsection{Overview}
\begin{figure}
    \centering
    \includegraphics[width=\linewidth]{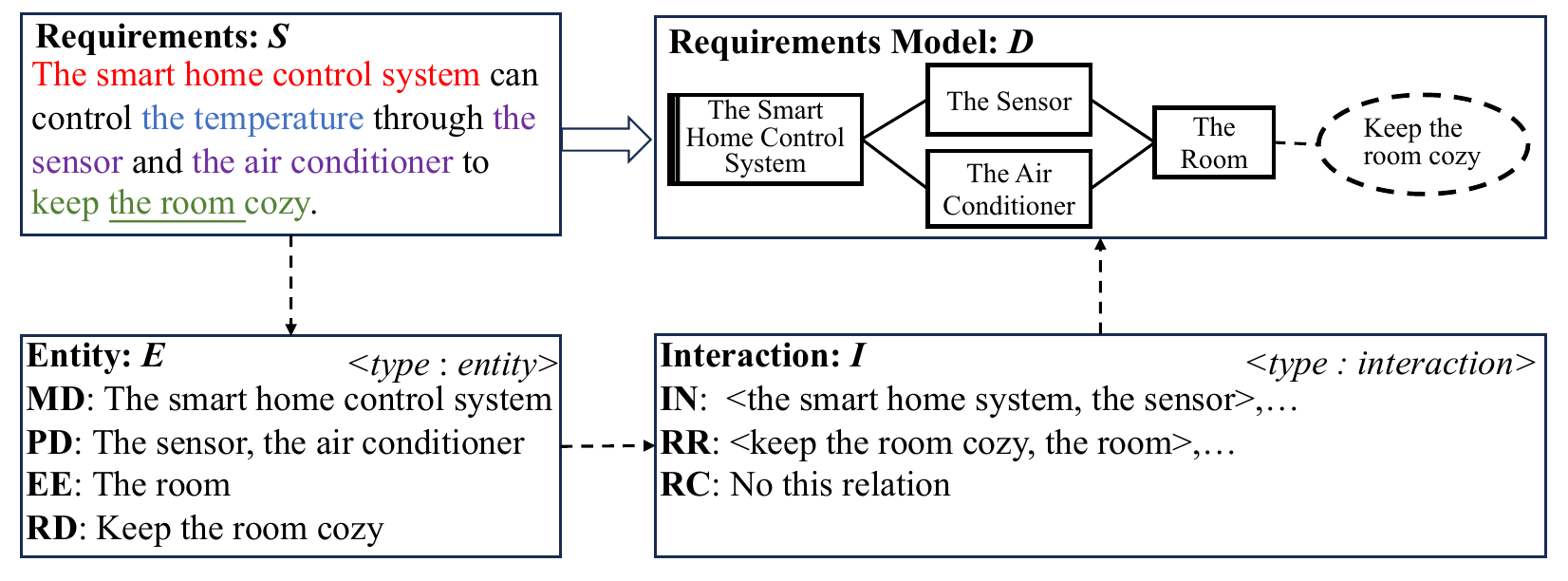}
    \caption{The formulation of CPSs requirements modeling}
    \label{fig:taskdefinition}
\end{figure}

As shown in Figure \ref{fig:taskdefinition}, the goal of requirements modeling is to construct requirements models from natural language requirements documents. This process involves identifying multiple dimensions of information (\eg \textit{Physical Device} and \textit{Interface Interaction}). Inspired by the manual process of requirements modeling, we decompose the requirements modeling into two types of identification tasks, i.e. the entity recognition and the interaction extraction. These tasks work in a pipeline as shown in Figure \ref{fig:taskdefinition}.

\subsection{Entity Recognition}
Given a natural language requirements description $S$ with length $M$, the entity recognition task is to identify entity elements $E = \{(e_i,t_i)\}_{i=0}^{N}$ contained in the requirements description, where $N$ is the number of entities and $e_i$, $t_i$ denote the value and type of the $i$-th entity, respectively.  
For CPSs requirements modeling with problem diagrams, there are six types of entities.
Detailed definitions of these types can be found in our annotation guidelines~\cite{web:code}. Below is a brief introduction to these types and their meanings.

\begin{itemize}
    \item Machine Domain (\textbf{MD}): is the software system that we want to build, such as \textit{the smart home control system}.
    \item Physical Device (\textbf{PD}): is the real-world device, which can be used to send or receive data, such as \textit{the sensor} and \textit{the air conditioner}.
    \item Environment Entity (\textbf{EE}): is the external object in the interactive environment, such as \textit{the user} and \textit{the operator}.
    \item Design Domain (\textbf{DD}): is the third-party system that already exists. Their properties are artificially designed or prescribed, such as \textit{database}.
    \item Requirements Domain (\textbf{RD}): is the purpose of the system to be developed, such as \textit{control the home environment}.
    \item Shared Phenomena (\textbf{SP}): is a set of shared events, states, and values between the connected entities, such as \textit{close notification} and \textit{click the button}.
\end{itemize}

\subsection{Interaction Extraction}
Given a natural language requirements description $S$ and $N$ entities $E = \{(e_i,t_i)\}_{i=0}^{N}$ recognized from the requirements description, the interaction extraction task is to find interactions $I = \{(h_j,r_j,t_j)\}_{j=0}^{M}$ among recognized entity $E$, where $h_j$ and $t_j$ is the head and tail entity of the $j$-th interaction. $r_i$ is the type of the $j$-th interaction.
For CPSs requirements modeling using problem diagrams, there are three types of interactions. Here are the types and their meanings.
\begin{itemize}
    \item Interface (\textbf{IN}): is the interface of shared phenomena between the connected entities, such as \textit{(the smart home control system, the notification)}.
    \item Requirements Reference (\textbf{RR}): is the reference interaction between requirements domain and other entities, such as \textit{(the patient, monitor the health condition)}. 
    \item Requirements Constraint (\textbf{RC}): is a constrain interaction between the requirements domain and other entities, such as \textit{(the medical watch, to monitor patient)}. It means the requirements domains do not just refer to the phenomena but constrain them.
\end{itemize}

\section{Benchmark Construction} \label{sec:benchmark}

\begin{figure*}
    \centering
    \includegraphics[width=\linewidth]{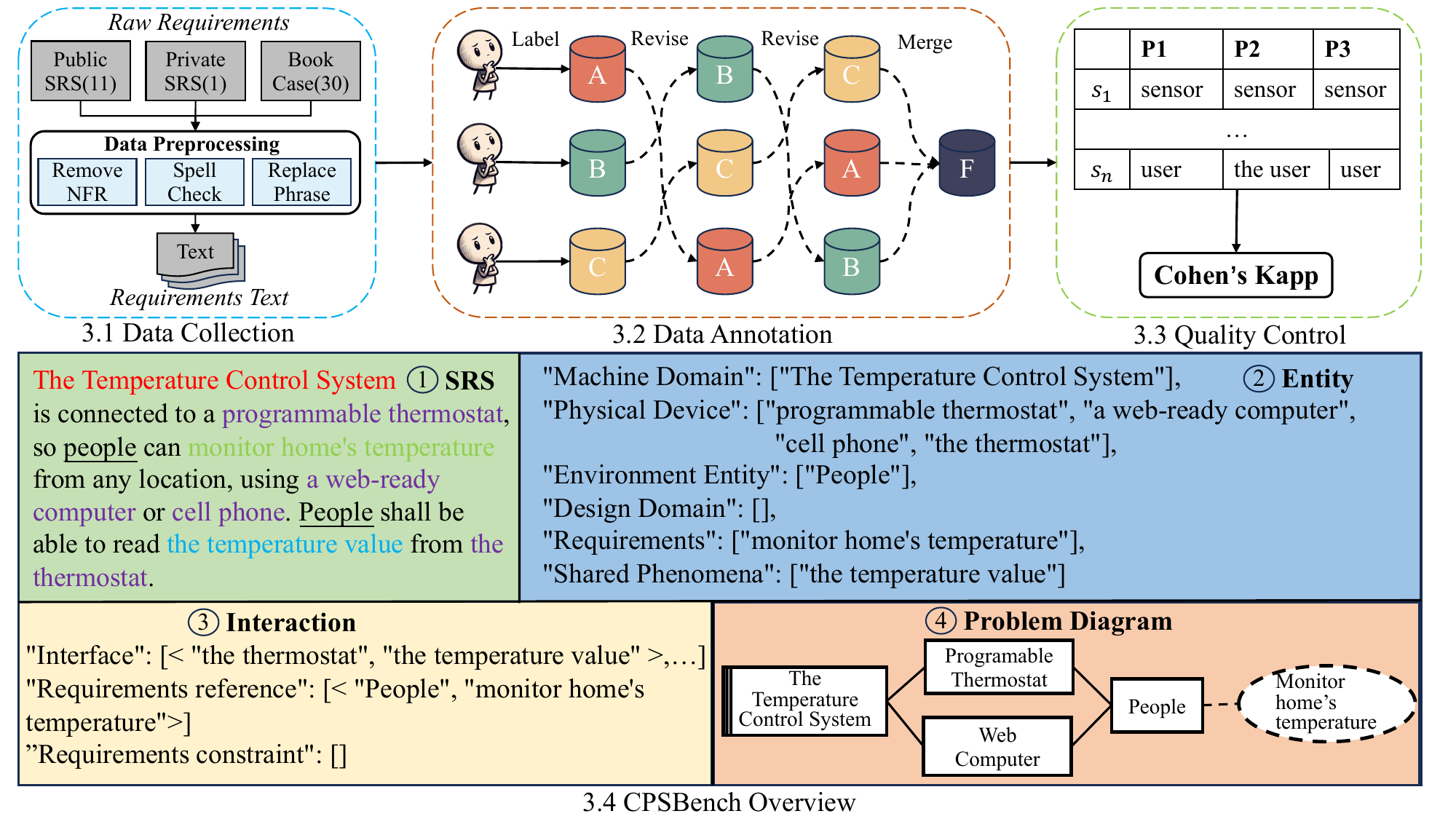}
    \caption{The Overview of {\sc CPSBench} Construction}
    \label{fig:pdd}
\end{figure*}

In this section, we propose a \textbf{CPS}s requirements modeling \textbf{Bench}mark, named {\sc CPSBench}. Figure \ref{fig:pdd} illustrates the process of constructing this benchmark. We describe the details in the following sections, including three steps and an example from the benchmark.

\subsection{Data Collection}

\textbf{Studies Systems.} Our study aims to evaluate the abilities of LLMs in requirements modeling for CPSs. To achieve this goal, we collect CPSs requirements documents covering diverse application domains, including embedded systems, control systems, and real-time systems. The documents originate from three sources: public software requirements documents datasets (\ie PURE~\cite{ferrari2017pure} and Lockheed Martin~\cite{mavridou2020ten}), private requirements documents (\ie SSCS) from industry~\cite{yang2022intelligent}, and cases from RE books~\cite{jackson2005problem}. In total, the collected CPSs requirements documents include: 
\begin{itemize}
    \item The Crime Tracking Network and Systems (CTS).
    \item Mars Expression Mission Ground Data System (MEM).
    \item The Space Fraction System (SFS).
    \item The Tactical Control System (TCS).
    \item The Correlator Monitor Control System (CCS).
    \item The Smart Home Control System (HCS).
    \item The Gemini Control System (GCS).
    \item Reversible Lane Control System (LCS).
    \item Telescope Control Flight System (FCS).
    \item Center-to-Center Network System (C2C).
    \item The Autopilot Control System (ACS).
    \item The Sun Search Control System (SSCS).
    \item Cases from books (Case).
\end{itemize}

PURE contains 79 software requirements documents from different domains. We manually reviewed them and selected 10 documents for CPSs. Lockheed Martin contains ten requirements documents from the cyber-physical domain. We reviewed these requirements documents and selected an autopilot control system for our benchmark. The criteria for selecting this document are project scale and areas of applicability. Specifically, the other 9 requirements documents are only about a page long. The Sun Search Control System (SSCS) is a private software requirements document from the aerospace domain. This SSCS has been used in multiple research works for evaluation~\cite{projection}~\cite{chatmodeler}. Additionally, we collected 30 cases from requirements engineering books~\cite{jackson2005problem}. 
 
\textbf{Preprocessing requirements documents}. The original software requirements documents often contain tables, figures, and incomplete sentences. To ensure the quality of the requirements documents, we clean and sample the original requirements documents from the following four aspects: (1) Remove the NFR: we first remove the catalog, titles, diagrams, and tables using the regex tool. 
At present, we limit our concerns to the functional requirements. Therefore, we manually review the requirements documents and remove non-functional requirements (NFR). We also remove sentences with no more than 10 words using the regex tool, as these sentences tend to be noisy data that do not contain requirements. (2) Spell Check: We perform a thorough spell check on the remaining requirements descriptions to ensure any typographical errors are corrected. We also rewrite incomplete sentences by hand because these sentences may create ambiguity and we focus on requirements modeling instead of requirements disambiguation. (3) Replace Phrase: We replace specific phrases and terminologies that are either ambiguous or inconsistent with standardized terminologies. For instance, we replace terms (\eg \textit{UAV}) with more specific phrases (\eg \textit{Unmanned Aerial Vehicle}). (4) We split requirements documents into sentences using Spacy tools~\cite{vasiliev2020natural}.

\subsection{Data Annotation}
\textbf{Ground-truth Labeling.} We use a web tool named \textit{label studio}~\cite{label} for the annotation process. We first provided annotators with the annotation guidelines~\cite{web:code} and conducted three meetings to learn about problem diagrams and the annotation tool. During the annotation process, we published the requirements descriptions in the tool. For each sentence, the annotators first manually label the modeling entities. Then, for each entity pair, the annotators judge whether an interaction exists and label its type. The labeled results are used as the ground truth for evaluation. To guarantee the correctness of the labeling results, we built an inspection team, which consisted of two PhD candidates and four master students majoring in computer science. All of them are fluent English speakers and have either conducted some research on requirements modeling or completed a semester course on requirements modeling. We divided the team into three groups and each group is responsible for four software systems (A, B, or C in Figure \ref{fig:pdd}). The labeled results from one group were reviewed by another group. When a labeled result received different opinions, we hosted a discussion with all team members to decide through voting. In total, we collected 2633 requirements description sentences from requirements documents and spent over 500 person-hours annotating 5795 entities and 3092 interactions. Table \ref{tab:statistics} provides a summary of the statistics for our {\sc CPSBench}. In Table \ref{tab:statistics}, ``+'' represents publicly available, ``-'' stands for private, and ``\#'' denotes the tutorial case.

\subsection{Quality Evaluation}
\textbf{Consistency Control.} To ensure consistency and high quality, we conducted a training phase for all annotators after course learning. At this stage, the six annotators were given a piece of requirements at a time to perform all annotation tasks. We then calculated the inter-annotator agreement (IAA)~\cite{artstein2017inter} between annotators using Cohen’s Kappa~\cite{mchugh2012interrater}, followed by disagreement discussion and guideline refinement. This process was repeated until the IAA score achieved \textit{substantial agreement} (\ie the IAA score is above 0.6)~\cite{mchugh2012interrater} . Afterward, the remaining set of requirements text was given to the annotators for annotation. The final Cohen's Kappa for labeling entities is 0.74, and the average Cohen's Kappa for labeling interaction is 0.78. These results demonstrate the quality and reliability of the annotations in {\sc CPSBench}.

\subsection{{\sc CPSBench} benchmark}
Figure \ref{fig:pdd} shows a sample in {\sc CPSBench}. Each sample consists of four components. \textbf{\ding{182} Requirements}: an English text description detailing the functional requirements of a software system. \textbf{\ding{183} Entity}: a dictionary that contains all modeling entities in requirements and their types. The key is the type of these entities, and the value is the name of these entities. \textbf{\ding{184} Interaction}: a dictionary that contains interactions among all modeling entities. \textbf{\ding{185} Problem Diagram}: a constructed requirements model with problem diagrams based on these entities and interactions.

\begin{table}[]
    \centering
    \caption{The statistics of the {\sc CPSBench} benchmark.}
    \begin{tabular}{cccccccccc}
\toprule
\multirow{2}{*}{\textbf{Sys}} & \multicolumn{6}{c}{\textbf{Entities}} & \multicolumn{3}{c}{\textbf{Interactions}} \\ \cline{2-10} 
                     & MD  & PD & EE & DD & RD & SP & IN       & RR       & RC      \\ \midrule
CTS+                 & 100 & 6  & 118& 20 & 82 & 109& 179      & 52       & 25      \\
MEM+                 & 37  & 17 & 60 & 21 & 22 & 55 & 127      & 4        & 14      \\
SFS+                 & 40  & 6  & 106& 5  & 18 & 7  & 11       & 2        & 6       \\
TCS+                 & 554 & 181& 281& 228& 376& 305& 536      & 140      & 35      \\
CCS+                 & 60  & 57 & 38 & 23 & 35 & 69 & 158      & 30       & 7       \\
HCS+                 & 58  & 118& 73 & 25 & 35 & 81 & 242      & 18       & 25      \\
GCS+                 & 71  & 132& 83 & 43 & 56 & 90 & 278      & 32       & 37      \\
LCS+                 & 43  & 72 & 65 & 33 & 47 & 76 & 153      & 42       & 27      \\
FCS+                 & 28  & 32 & 37 & 56 & 38 & 86 & 152      & 54       & 33      \\
C2C+                 & 89  & 76 & 91 & 24 & 17 & 227& 203      & 14       & 26       \\ 
ACS+                 & 27  & 31 & 37 & 41 & 19 & 73 & 67       & 21       & 17      \\\midrule
SSCS-                & 31  & 43 & 78 & 56 & 28 & 172& 213      & 42       & 33       \\ \midrule
Case\#               & 52  & 63 & 68 & 71 & 32 & 63 & 77       & 31       & 34        \\ \midrule
\textbf{Total}  &1190 & 834 &1135 &646    &805 &1413      &2386      &483          &319         \\ 
\bottomrule
\end{tabular}
    \label{tab:statistics}
\end{table} 
\section{Approach} \label{sec:approach}
In this section, we describe the approach for evaluating the performance of LLMs in modeling CPSs requirements with the problem diagram. We formally define the overview of the approach and describe the details in the following sections, including prompt construction and few-shot retrieval, response generation, and answer parsing.

\subsection{Overview}
Our approach follows the general paradigm of in-context learning (ICL) and can be decomposed into three steps: 
\begin{itemize}
    \item \textbf{Prompt Construction}: for a given requirements text, we retrieve $k$ shots and construct a prompt with them to instruct the LLMs.
    \item \textbf{Response Generation}: feed the constructed prompt to LLMs to obtain the response sequence.
    \item \textbf{Answer Parsing}: transform the response sequence to a list of entities and interactions that can be assembled into a requirements model.
\end{itemize}

\subsection{Prompt Construction}
\begin{figure}
    \centering
    \includegraphics[width=1.0\linewidth]{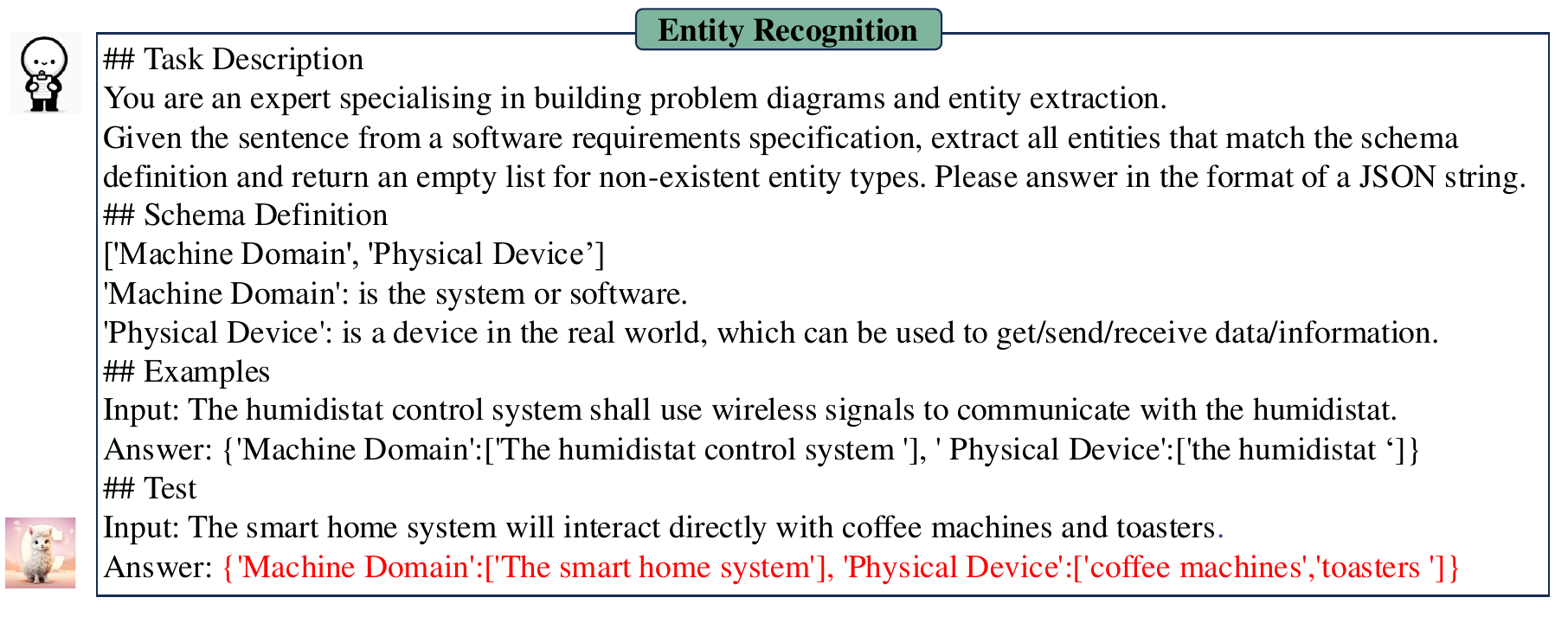}
    \caption{An example of the prompt of entity recognition}
    \label{fig:approach-e}
\end{figure}

\begin{figure}
    \centering
    \includegraphics[width=1.0\linewidth]{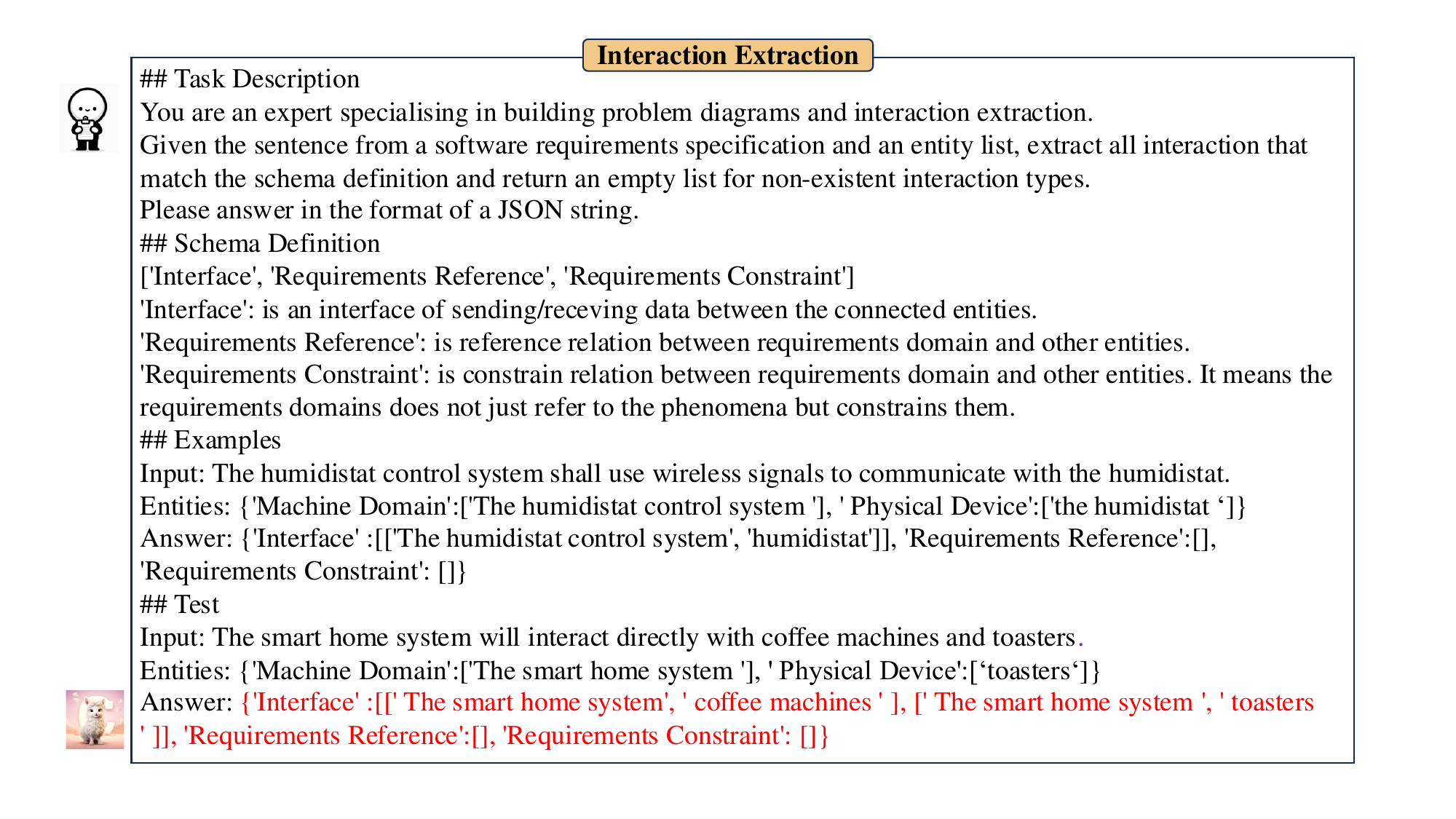}
    \caption{An example of the prompt of interaction extraction}
    \label{fig:approach-r}
\end{figure}

Figure \ref{fig:approach-e} and Figure \ref{fig:approach-r} show the constructed prompts for the entity recognition task and interaction extraction task, respectively. The prompt $P$ consists of four parts: task description $T$, schema definition $S$, few-shot examples $E$, and requirements text for test $R$.

\begin{equation}
    P = (T, S, E, R)
\end{equation}

\textbf{Task Description ($T$)}: gives an overview of the task. The first sentence of the task description \textit{``You are an expert...''} instructs LLMs to use the knowledge of CPSs requirements modeling and information extraction. The second sentence \textit{``Given the sentence from a software requirements specifications...''} makes LLMs understand the task of entities or interactions recognition. The last sentence informs LLMs about the output format for easy parsing.

\textbf{Schema Definition ($S$).} Schema definition details all target entity types or interaction types in the requirements model, along with their definition based on prior work~\cite{abs-2404-00152}.

\textbf{Few-shot Examples ($E$).} The few-shot examples are appended to the prompt for two purposes: (1) to provide the LLMs with tutorials and references about the modeling task for making predictions, and (2) to regulate the format of the LLMs output for each input, as LLMs will generate output that mimics the format of shots. This is crucial for parsing the generated output to get the final requirements models.

\textbf{Requirements Text ($R$).} This part feeds the CPSs requirements text into the LLMs and expects them to identify all entities or interactions based on the defined schema.

\subsection{Few-shot Retrieval}
Inspired by prior work~\cite{wang2023gpt}, we use semantic similarity retrieval to obtain $k$ shots. This involves retrieving $k$ shots with close semantics from the training set $C$ for each input test requirements text. Specifically, we first use text similarity models $Encoder$ to compute the embedding $H = \{h_i\}_{i=1}^{n}$ for all training samples $C = \{c_i\}_{i=1}^n$ and the embedding $h'$ for input requirements text $R$, where $n$ is the number of samples in the training set, $c_i$ and $h_i$ denote the requirements description and embedding of $i$-th sample in the training set, respectively.

\begin{equation}
\begin{aligned}
    H &= \operatorname{Encoder}(C)\\
    h' &= \operatorname{Encoder}(R) \\
\end{aligned}
\end{equation}

Then, we use cosine similarity to compute the embedding similarity $m_i$ between the embedding of $i$-th sample $h_i$ and the embedding of input requirements $h'$. Thus, we can get all similarity scores $M = \{m_i\}_{i=1}^{n}$.

\begin{equation}
\begin{aligned}
    m_i &= \frac{h_i \cdot h'}{|h_i| \cdot |h'|} 
\end{aligned}
\end{equation}

Finally, we sort each sample in descending order based on the embedding similarity score $M$ and get the top $k$ shots $e$.
\begin{equation}
\begin{aligned}
    e &= \operatorname{top}(\operatorname{sort}(M))
\end{aligned}
\end{equation}

\subsection{Response Generation}
After retrieving the k-shots, we get the constructed prompt $P$. Then we feed the prompt $P$ to LLMs to generate response $G$. 
We choose greedy sampling~\cite{wu2019active} with temperature $t$ as 0. Greedy sampling is chosen to avoid the randomness associated with other sampling methods, ensuring consistency and reliability in the generated answers.


\subsection{Answer Parsing}
Since LLMs may not always generate a response in the exact format specified in the prompt, we use regular expression to parse the generated response $G$ and format it into answer $A$ for easier metric computation. Specifically, we search for content within \textit{'\{\}'} and extract the needed part.


\section{study design} \label{sec:studydesign}
To evaluate the performance of LLMs in CPSs requirements modeling from natural language documents, we conduct a large-scale study to answer three research questions. In this section, we describe the details of our study, including research questions, studied LLMs, metrics, and experiment setting.

\subsection{Research Questions}
Our study aims to answer three research questions (RQs). In RQ1, we evaluate the performance of LLMs in recognizing entities and extracting interactions from CPSs requirements documents. In RQ2, we conduct experiments to estimate the effect of the number of shots in requirements modeling. In RQ3, we investigate and summarize the type of hallucinations in CPSs requirements modeling with LLMs.

\textbf{RQ1: What is the performance of LLMs in entities and interactions recognition from CPSs requirements documents?} We use 10-fold cross-validation to divide the {\sc CPSBench} into training and test datasets. Then we retrieve shots from the training dataset for each sample in the test dataset to construct the prompt. Last, we feed the prompt to LLMs and use multiple metrics to evaluate the performance of LLMs in entity and interaction recognition.

\textbf{RQ2: How does the number of shots affect the performance of LLMs in CPSs requirements modeling?} We also first split the {\sc CPSBench} into training and test datasets. Then, we retrieve different numbers of shots to estimate the impact on requirements modeling. Given the limitations of response speed, the number of shots ranges from 1 to 3.

\textbf{RQ3: What are the hallucinations in requirements modeling with LLMs?} To further enhance the ability of LLMs, we take gpt-4 for hallucination analysis. We manually reviewed the ground truth and gpt-4 predictions for each test sample and summarized the statistics of hallucination types.  

\begin{table}[]
    \centering
    \caption{Evaluated LLMs in our benchmark}
    \begin{tabular}{ccccc}
\toprule
\textbf{Type}                 & \textbf{Name}   & \textbf{Version}  & \textbf{Context}  & \textbf{Publisher} \\ \midrule
\multirow{2}{*}{Close-source} & gpt-4   & gpt-4-turbo-0409   & 128,000  & OpenAI\\
                              & gpt-3.5 & gpt-3.5-turbo-0125 & 16,385   & OpenAI\\ \midrule
\multirow{5}{*}{Open-source}  & Qwen 2  & 7B                 & 128,000  & Alibaba\\
                              & LLama3  & 8B                 & 8,192    & Meta AI\\
                              & Gemma2   & 7B                 & 8,192    & Google\\
                              & glm4    & 9B                 & 8,192    & THUDM\\
                              & Mistral & 7B                 & 8,192    & Mistral AI\\ \bottomrule
\end{tabular}
    \label{tab:llms}
\end{table}

\begin{table*}[]
    \centering
    \caption{The results of LLMs on requirements modeling for CPSs}
    \begin{tabular}{cccccccccccc}
\toprule
                       & \multicolumn{7}{c}{\textbf{Entity}}                                                 & \multicolumn{4}{c}{\textbf{Interaction}}                       \\ \cline{2-12} 
\multirow{-2}{*}{\textbf{LLMs}} & MD    & PD    & EE    & DD    & RD    & SP    & {\color[HTML]{FD6864} Ave} & IN    & RR    & RC    & {\color[HTML]{FD6864} Ave} \\ \midrule
\multicolumn{12}{c}{1-shot}                                                                                                                              \\ \midrule
gpt-4                  & 54/75/63 & 45/48/46 & \textbf{49/41/45} & 53/27/36 & 15/39/22 & 37/41/40 & 38/48/43 & \textbf{72/59/65} & 24/35/29 & 24/44/31 & \textbf{59/55/57}             \\
gpt-3.5                & \textbf{66/71/68} & \textbf{56/56/56} & 40/40/41 & 30/35/32 & \textbf{22/37/28} & 34/41/37 & \textbf{41/49/45}& 72/55/63 & 26/30/28 & 17/44/25 & 58/52/55          \\
Qwen2                  & 60/65/63 & 50/33/40 & 49/41/45 & 33/27/30 & 19/30/23 & \textbf{38/46/42} & 42/44/43 & 51/51/51 & 50/30/38 & 15/44/23 & 47/49/48          \\
LLama3                 & 48/59/53 & 56/33/42 & 35/35/35 & \textbf{46/32/38*} & 21/33/26 & 44/32/37 & 40/39/40 & 55/57/56 & 28/39/33 & \textbf{27/44/33} & 50/54/52          \\
Gemma2                 & 49/69/57 & 44/54/48 & 44/41/42 & 17/30/22 & 10/28/15 & 39/45/42 & 32/47/38 & 56/55/55 & 20/52/29 &  8/44/14 & 39/54/46          \\
glm4                   & 67/67/67 & 50/46/48 & 43/38/41 & 24/22/23 & 13/31/19 & 42/41/42 & 38/44/41 & 72/42/53 & 31/26/27 & 27/33/30 & 63/40/49          \\
Mistral                & 63/70/66 & 43/28/34 & 44/32/37 & 24/22/23 & 19/30/23 & 42/38/40 & 41/40/41 & 60/50/54 & \textbf{45/43/44} & 21/44/29 & 54/49/51          \\ \midrule
\multicolumn{12}{c}{2-shot}                                                                                                                              \\ \midrule
gpt-4                  & 56/79/66 & 57/54/55 & \textbf{58/49/53*} & \textbf{50/27/35} & 19/46/27 & \textbf{39/53/45*} & \textbf{43/55/48*} & \textbf{74/77/74} & 40/52/45 & \textbf{31/44/36*} & \textbf{68/68/68}          \\
gpt-3.5                & 60/78/68 & 45/46/45 & 45/54/49 & 29/27/28 & 22/37/27 & 31/36/37 & 40/52/45 & 66/52/58 & 67/35/46 & 14/44/22 & 58/50/53          \\
Qwen2                  & 75/51/60 & 54/13/21 & 53/26/35 & 17/16/16 & 25/20/22 & 40/28/33 & 45/29/35 & 63/60/61 & \textbf{44/52/48*} & 19/44/27 & 56/58/57          \\
LLama3                 & 54/62/58 & 67/44/53 & 44/41/42 & 38/30/33 & 25/43/22 & 41/38/39 & 40/39/40 & 54/58/56 & 30/43/36 & 27/44/33 & 49/55/52          \\
Gemma2                 & 51/76/61 & 45/54/49 & 48/49/48 & 21/41/27 & 14/33/20 & 37/45/41 & 35/52/42 & 55/62/58 & 25/61/35 & 8/44/13 & 41/61/49                      \\
glm4                   & 64/76/70 & \textbf{62/59/60*} & 47/40/43 & 29/35/32 & \textbf{20/46/40*} & 29/33/31 & 40/50/45 & 66/55/60 & 31/48/37 & 16/44/24 & 53/53/53          \\
Mistral                & \textbf{66/73/70} & 53/33/41 & 51/37/43 & 24/27/25 & 28/48/35 & 35/34/35 & 43/45/44 & 63/55/59 & 42/43/43 & 24/44/31 & 58/53/55          \\ \midrule
\multicolumn{12}{c}{3-shot}                                                                                                                              \\ \midrule
gpt-4                  & \textbf{63/83/72*} & \textbf{63/57/60*} & \textbf{55/49/52} & \textbf{50/30/37} & 16/37/22 & 34/49/40 & \textbf{43/54/48*} & \textbf{77/73/75*} & \textbf{37/48/42} & \textbf{31/44/36*} & \textbf{68/69/69*}          \\
gpt-3.5                & 61/75/68 & 40/43/41 & 46/47/46 & 36/24/29 & 25/46/32 & 25/39/31 & 39/49/43 & 68/63/65 & 35/31/33 & 14/44/22 & 57/58/58          \\
Qwen2                  & 58/11/18 & 63/9/16 & 50/10/17 & 33/3/5 & 30/6/16 & 40/13/20 & 47/10/16     & 67/64/66 & 36/43/39 & 19/44/27 & 58/61/60          \\
LLama3                 & 52/60/55 & 61/41/49 & 40/43/41 & 41/24/31 & 23/35/28 & 32/29/31 & 43/45/44 & 55/62/59 & 23/39/29 & 25/67/36 & 47/60/53                      \\
Gemma2                 & 52/76/62 & 45/56/50 & 47/51/49 & 19/32/24 & 15/33/21 & 32/50/39 & 35/53/42 & 53/62/57 & 23/43/30 & 8/44/14 & 41/59/48                      \\
glm4                   & 62/73/67 & 54/50/52 & 47/43/45 & 31/30/30 & 18/37/23 & 30/41/35 & 39/48/43 & 71/59/64 & 34/48/40 & 10/36/15 & 56/56/56          \\
Mistral                & 65/71/68 & 56/37/44 & 54/43/48 & 33/27/30 &\textbf{32/48/38} & \textbf{43/45/44} & 48/48/48 & 70/60/65 & 30/35/32 & 21/44/29 & 60/57/58          \\ \bottomrule
\end{tabular}
    \label{tab:rq1}
\end{table*}

\subsection{Studied LLMs} \label{sec:llms}
Table \ref{tab:llms} shows 7 evaluated LLMs in our experiments. They are the latest versions of the LLMs released by well-known companies or organizations. They cover closed-source LLMs (i.e., gpt-4~\cite{openai2023gpt}, gpt-3.5~\cite{gpt-3.5}) and open-source LLMs (i.e. Qwen2~\cite{Qwen}, LLama3~\cite{LLama3}, Gemma~\cite{gemma}, glm4~\cite{glm} and Mistral\cite{Mistral-7B}). We use official interfaces or implementations to reproduce these LLMs.

\subsection{Evaluation Metrics} \label{sec:metrics}
Following previous studies~\cite{shen2023promptner}~\cite{wang2022matchprompt}, we use micro precision, recall, and F1 score to assess the effectiveness of entities and interactions recognition. Specifically, we first compute the count of correctly identified entities or interactions (TP), the count of entities or interactions identified by LLMs but not present in the gold standard (FP), and the count of entities or interactions in the gold standard but not identified by LLMs (FN). Then we aggregate the TP, FP, and FN across all entities or interactions and compute precision (P) and recall (R). Last we compute the F1 score based on precision and recall.

\begin{equation}
\begin{aligned}
    P &= \frac{TP}{TP + FP} \\
    R &= \frac{TP}{TP + FN} \\
    F1 &= 2 \cdot \frac{P \cdot R}{P + R}
\end{aligned}
\end{equation}

\subsection{Experiment Settings}
The experiment settings of our evaluation are as follows:

\textbf{Dataset Split.} To make sure that all requirements text are only used either in the training dataset or test dataset, we conduct the \textbf{10-fold cross validation} on the sentence level. Specifically, we first randomly partition all the requirements text into 10 parts. One single part is retained as the testing data, and the rest 9 parts are used as the training data. Then we repeat 10 times.

\textbf{LLMs Setting.} For close-source LLMs, we implement gpt-3.5 and gpt-4 by invoking OpenAI's API~\cite{openaiapi}. For open-source LLMs, such as Qwen2 and LLama3, we instantiate them with their replication packages and download their pre-trained weight from HuggingFace. The default settings of LLMs are the same, using greedy sampling\cite{chickering2002optimal} with \textit{temperature}=0. We also use the same inference library - vllm~\cite{web:vllm} for LLMs inference and serving to ensure the fairness of our evaluation.


\textbf{Shot Retrieval.} We first use the open-source framework - SimCSE~\cite{gao2021simcse} and pre-trained model - \textit{princeton-nlp/sup-simcse-bert-base-uncased} to compute embeddings for all training examples and the test requirements text. Then, we use faiss~\cite{douze2024faiss} to build the index and get the top 3 similar samples from the training set.
\section{result and analysis} \label{sec:resultandanalysis}
In our first research question, we evaluate the performance of LLMs in CPSs requirements modeling from requirements documents, including entity and interaction identification.

\textbf{RQ1: What is the performance of LLMs in entities and interactions recognition from requirements documents?}

\textbf{Setup.} We evaluate advanced LLMs (Section \ref{sec:llms}) on our constructed our {\sc CPSBench} (Section \ref{sec:benchmark}) with 1-shot reasoning. The evaluation metrics are described in Section \ref{sec:metrics}, \ie the Precision (P), Recall (R), and F1 score. For all metrics, higher scores represent better performance. 

\begin{figure*}[h]
	
	\begin{minipage}{0.32\linewidth}
		\vspace{3pt}
		\centerline{\includegraphics[width=\textwidth]{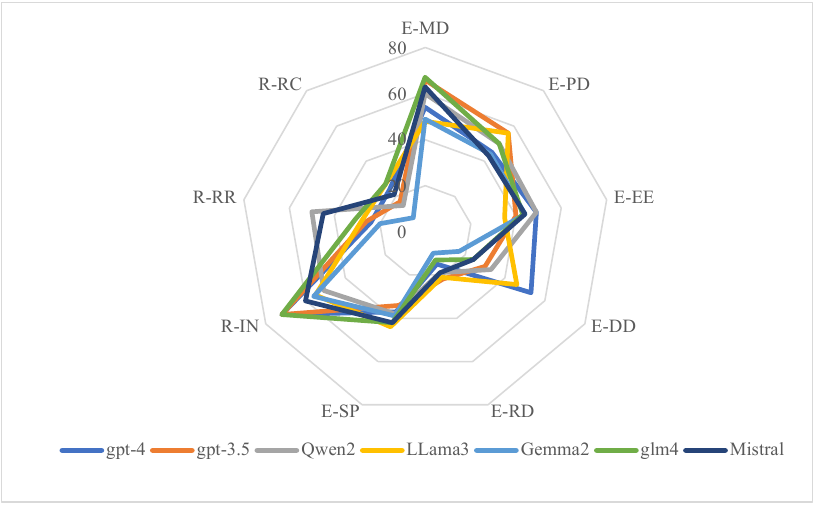}}
		\centerline{\small The precision score of LLMs}
	\end{minipage}
	\begin{minipage}{0.32\linewidth}
		\vspace{3pt}
		\centerline{\includegraphics[width=\textwidth]{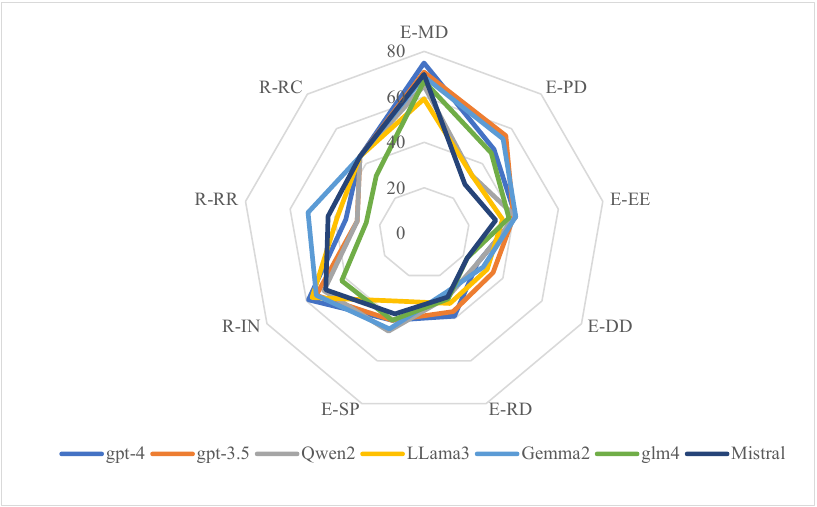}}
		\centerline{\small The recall score of LLMs}
	\end{minipage}
	\begin{minipage}{0.32\linewidth}
		\vspace{3pt}
		\centerline{\includegraphics[width=\textwidth]{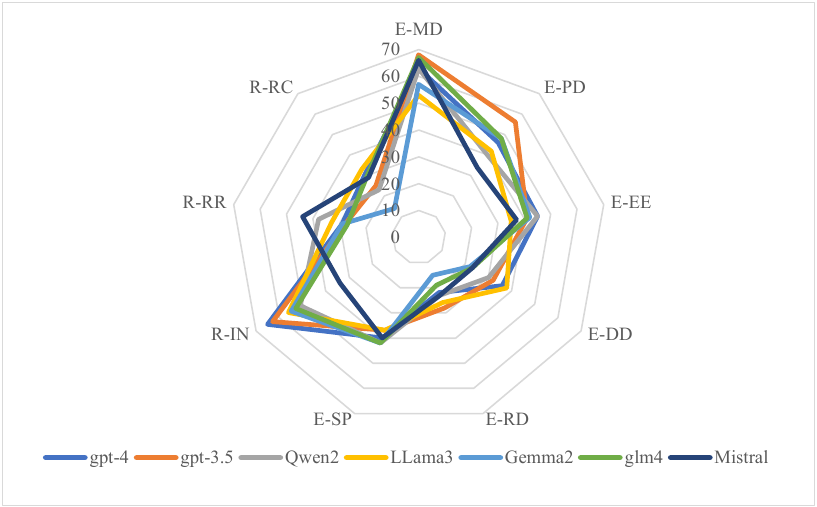}}
		\centerline{\small The f1 score of LLMs}
	\end{minipage}
	\caption{Visual comparisons of LLMs in problem diagram extraction.}
	\label{fig:rq1}
\end{figure*}

\textbf{Results.} Table \ref{tab:rq1} and Figure \ref{fig:rq1} show the experimental results of 7 LLMs on the {\sc CPSBench}. Each cell in the Table \ref{tab:rq1} contains three numbers representing P, R, and F1. Besides, in Table \ref{tab:rq1}, each bolding represents the best performance with the same number of shots. ``*'' represents the best performance of all results. In Figure \ref{fig:rq1}, ``E-'' represents the entity, and ``R-'' represents the interaction in problem diagrams.


\textbf{Analyses.} \uline{(1) The ability of LLMs with few-shot reasoning to model CPSs requirements from requirements documents is limited.} The average recall and f1 score of LLMs are only about 60 (the Ave column in Table \ref{tab:rq1}). Since the purpose of using LLMs to model CPSs requirements is to save requirements engineers time in reading requirements documents, the recall rate is more valued. A higher recall rate means fewer entities or interactions are not identified. However, almost half of the entities or interactions are not identified correctly by LLMs. On the one hand, we believe this is due to the CPSs requirements containing domain-specific knowledge of physical components, which is difficult to understand. On the other hand, the problem frame contains many types of elements, making identification prone to errors. We provide an error analysis in the following RQ3. \uline{(2) gpt-3.5 and gpt-4 achieve the best results among all LLMs for entity and interaction recognition, respectively.} For entity identification, the average f1 score of gpt-3.5 with one-shot reasoning achieves 45. Compared to the other LLMs, gpt-3.5 outperforms them from 4\% to 18.4\%. For interaction identification, gpt-4 achieves 57 and outperforms other LLMs from 5\% to 19\%. \uline{(3) LLMs differ in their ability to understand different entities or interactions.} From Figure \ref{fig:rq1}, we can see that LLMs exhibit a better understanding of general requirements concepts, but their performance is relatively inferior on specialized concepts. Specifically, LLMs have a richer knowledge of the machine domain (E-MD), physical devices (E-PD), environmental entity (E-EE), and interface (R-IN). However, LLMs lack knowledge of concepts such as design domain (E-DD), requirements domain (E-RD), requirements reference (R-RR), and requirements constraints (R-RC). We believe this is because there is not much material related to these specialized concepts when LLMs are trained. 

\begin{center}
\fcolorbox{black}{gray!10}{\parbox{.9\linewidth}{\textbf{Answer to RQ1:} Although LLMs using few-shot reasoning have limited capability in CPSs requirements modeling, gpt-3.5 and gpt-4 achieve the best performance in entity and interaction recognition, respectively. Besides, LLMs vary in their ability to understand different entities or interactions and lack knowledge of specialized concepts in CPSs modeling. }}
\end{center}

\textbf{RQ2: How does the number of shots affect the performance of LLMs in requirements modeling for CPSs?}  

\textbf{Setup.} In this RQ, we first retrieve the different numbers of shots and put them into the constructed prompt (as shown in Figure \ref{fig:approach-e} and Figure \ref{fig:approach-r}). Then we feed these prompts to the LLMs and evaluate these LLMs on {\sc CPSBench}. Given the context length of LLMs, we set the number of shots to range from 1 to 3. The evaluation metrics are also the precision, recall, and f1 score. 

\begin{figure}
    \centering
    \includegraphics[width=0.9\linewidth]{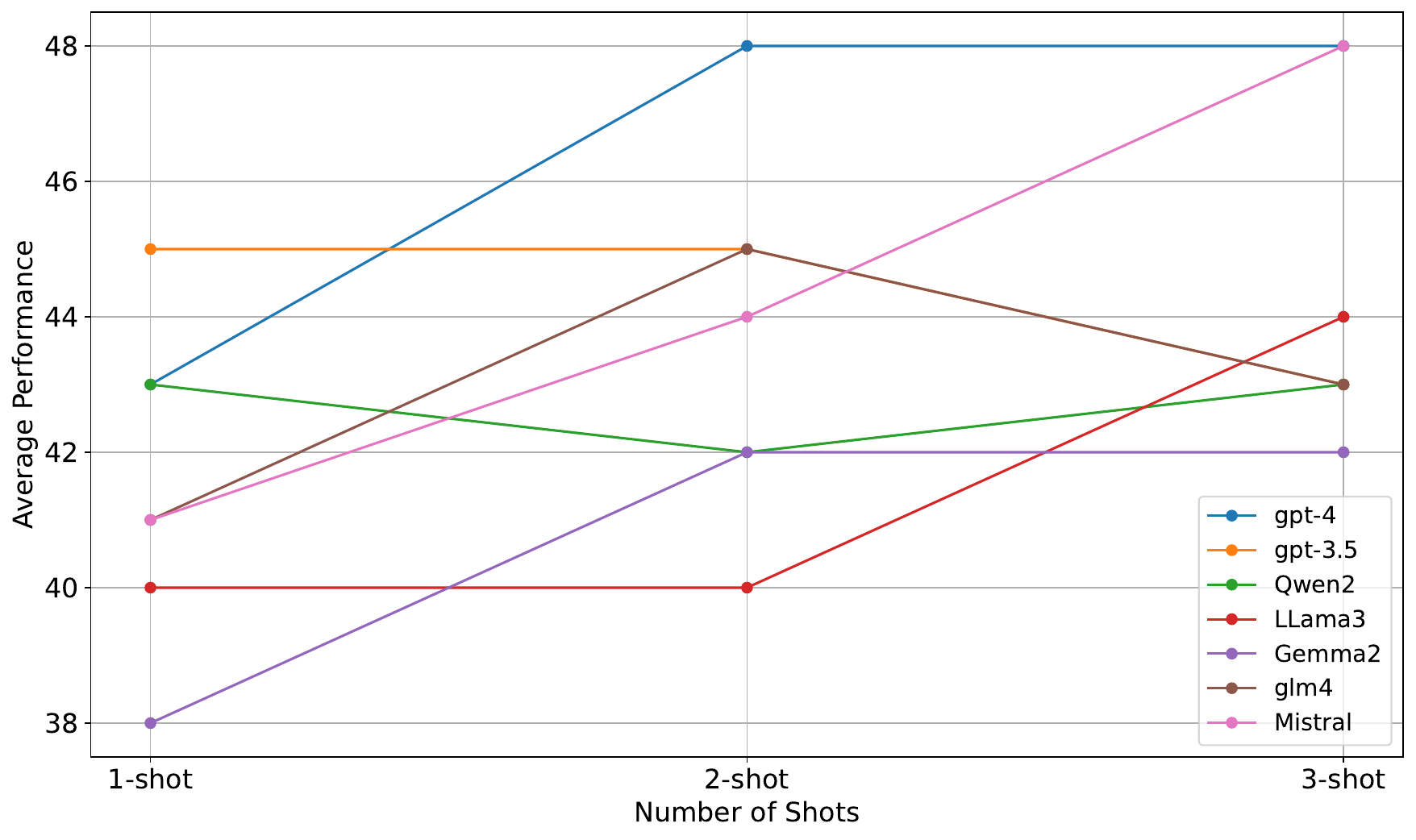}
    \caption{The performance of entity recognition by varying k-shot}
    \label{fig:rq2-e}
\end{figure}

\begin{figure}
    \centering
    \includegraphics[width=0.9\linewidth]{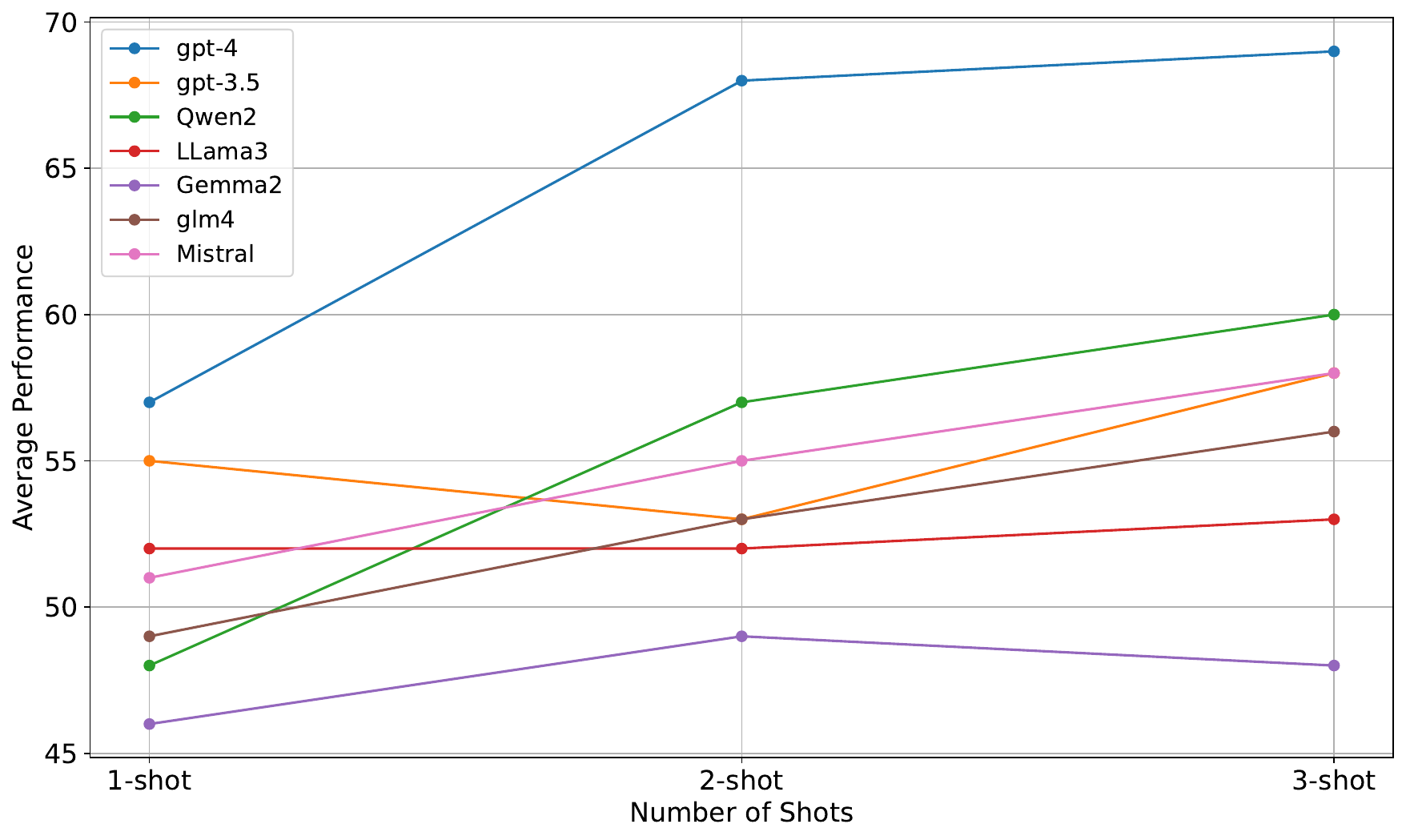}
    \caption{The performance of interaction extraction by varying k-shot}
    \label{fig:rq2-r}
\end{figure}

\textbf{Results.} The experimental results is shown in Table \ref{tab:rq1}. Figure \ref{fig:rq2-e} and Figure \ref{fig:rq2-r} visualize the f1 score of these advanced LLMs (Section \ref{sec:llms}) on entity recognition and interaction extraction from 1 to 3 shots, separately.

\begin{table*}[]
    \centering
    \caption{Analysis of LLMs hallucinations in requirements
modeling}
    \begin{tabular}{ccccp{10cm}}
    \toprule
    \bf Task & \multicolumn{2}{c}{\bf Hallucinations} & \bf Type & \multicolumn{1}{c}{\bf Case}  \\
    \midrule
    \multirow{18}{*}{\makecell{Entity \\ Recognition}} & \multicolumn{2}{c}{\multirow{3}{*}{Type Error}} & Input & The Tactical Control System will be capable of being hosted on computers. \\
                                        &                             & & Ground &  Physical Device:[computers] \\
                                        &                             & & Prediction & Design Domain:[computers] \\
    \cmidrule{2-5}
                                        & \multirow{9}{*}{\makecell{Boundary \\ Error}} & \makecell{Contain gold} & Input & Tactical Control System provide the capability to control the AV's Identification Friend. \\
                                        &                             & & Ground & Requirements:[control the AV's Identification Friend] \\
                                        &                             & & Prediction & Requirements:[provide the capability to control the AV's Identification Friend] \\
    \cmidrule{3-5}
                                        &                             & \makecell{Contained by gold} & Input & The thermostat shall allow a user to monitor and control a home’s temperature. \\
                                        &                             & & Ground & Requirements: [to monitor and control a home’s temperature] \\
                                        &                             & & Prediction & Requirements: [to monitor and control a home] \\
    \cmidrule{3-5}
                                        &                             & \makecell{Overlap with gold} & Input & A button providing an opportunity to explore content related to the thematic elements. \\
                                        &                             & & Ground & Requirements:[to explore content] \\
                                        &                             & & Prediction & Requirements:[explore content related to the thematic elements.] \\
    \cmidrule{2-5}
                                        & \multicolumn{2}{c}{\multirow{3}{*}{Complete Error}} & Input & A user shall be able to monitor and control home devices and systems. \\
                                        &                             & & Ground & Environment Entity:[a user]. \\
                                        &                             & & Prediction & Environment Entity:[home] \\
    \cmidrule{2-5}
                                        & \multicolumn{2}{c}{\multirow{3}{*}{Omitted Entities}} & Input & Tactical Control System software provide a windows based graphic operator interface. \\
                                        &                             & & Ground & Environment Entity:[operator] \\
                                        &                             & & Prediction & Environment Entity:[] \\
    \midrule
    \multirow{9}{*}{\makecell{Interaction \\ Extraction}} & \multicolumn{2}{c}{\multirow{3}{*}{Type Error}}& Input & The CCTNS system should run on multiple browsers. \\
                                        &                             & & Ground & Requirements Reference:[The CCTNS system, run on multiple browsers] \\
                                        &                             & & Prediction & Requirements Constraints:[The CCTNS system, run on multiple browsers] \\
    \cmidrule{2-5}
                                        & \multicolumn{2}{c}{\multirow{3}{*}{Complete Error}} & Input & The CMCS system performs limited amounts of real-time data to collect products. \\
                                        &                             & & Ground & Requirements Reference: [The CMCS system, to collect products] \\
                                        &                             & & Prediction & Requirements Reference: [to collect products, limited amounts of real-time data] \\
    \cmidrule{2-5}
                                        & \multicolumn{2}{c}{\multirow{3}{*}{Omitted Interactions}} & Input & The CCTNS system must be able to export audit trails for specified cases. \\
                                        &                             & & Ground & Requirements Reference:[export audit trails, The CCTNS system] \\
                                        &                             & & Prediction & Requirements Reference:[] \\
    \bottomrule
    \end{tabular}
    \label{tab:rq3}
\end{table*}

\textbf{Analyses.}  \uline{(1) The ability of LLMs to model CPSs requirements can improve with more shots.} As shown in Figure \ref{fig:rq2-e} and Figure \ref{fig:rq2-r}, the performance of LLMs in entity and interaction recognition continues to rise as the number of shots increases except Qwen2. However, as the number of shots increases, the magnitude of improvement will decrease. Specifically, the average performance of gpt-4 on entity recognition increased 12\% on the F1 score from 1 shot to 2 shots and kept the same score (\ie 48) from 2 shots to 3 shots. The average performance of gpt-4 on interaction extraction increased 16\% on the F1 score from 1 shot to 2 shots and only increased 1\% from 2 shots to 3 shots. Thus, providing more shots helps LLMs better understand the requirements context and the knowledge of the requirements model, leading to an improvement in performance. \uline{(2) gpt-4 is most sensitive to the effects of shot number.} From 1 shot to 2 shots, the entity recognition of gpt-4 increased 12\% on the f1 score while glm4 only increased 7\%. For interactions extraction, gpt-4 increased 16\% while glm4 only increased 8\%. This means that gpt-4 may be better at integrating and synthesizing knowledge from retrieved shots, which allows it to make its predictions more effectively when provided with extra shots.

\begin{center}
\fcolorbox{black}{gray!10}{\parbox{.9\linewidth}{\textbf{Answer to RQ2:} The number of shots has a substantial impact on the performance of LLMs in requirements modeling for CPSs. However, the benefits will diminish with further increases.}}
\end{center}

\textbf{RQ3: What are the hallucinations in requirements
modeling with LLMs?}

\textbf{Setup.} To further improve the ability of
LLMs, we conduct hallucination analysis for gpt-4. We first use an open-source package - gradio~\cite{gradio} to create a web application. This application shows the requirements text, ground truth, and prediction by gpt-4. Then We invited a PhD candidate and a master student to review these results and summarize the statistics of hallucination types. They are familiar with CPSs requirements modeling and problem diagrams.

\textbf{Results.} Table \ref{tab:rq3} summarizes the statistics of hallucinations and their corresponding cases. 

\textbf{Analyses.}  For entity recognition, it has four types of hallucinations. \uline{(1) Type Error}: errors on the type of an entity. \uline{(2) Boundary Error}: errors on the identification of boundaries, including \textit{contain gold}, \textit{contained by gold}, and \textit{overlap with gold}. The \textit{Contain gold} means the predicted entity contains the correct entity. The \textit{Contained by gold} means the predicted entity is included by the correct entity. The \textit{Overlap with gold} means that the predicted entity overlaps with the correct entity, but neither contains nor is contained. \uline{(3) Complete Error}: errors that are completely outside the gold entity. \uline{(4) Omitted Entities}: errors in the omission of entities from the prediction. 

There are three types of hallucinations for interaction extraction. \uline{(a) Type Error}: errors on the type of interactions between an entity pair.  \uline{(b) Complete Error}: errors that the prediction is completely different from the gold results. \uline{(c) Omitted Interactions}: errors in the omission of interactions. 

\begin{center}
\fcolorbox{black}{gray!10}{\parbox{.9\linewidth}{\textbf{Answer to RQ3: } We summarize 4 types of hallucinations on entity recognition, including type error, boundary error, complete error, and omitted entities. We also summarize 3 types of hallucinations on interaction extraction, including type error, complete error, and omitted interaction.}}. 
\end{center}




\section{Discussion} \label{sec:discussion}


\subsection{Threats to validity} 
\textbf{Construct Validity} concerns the relationship between the treatment and the outcome. The threat comes from the rationality of the research questions we posed. We are interested in assessing the effectiveness of LLMs in understanding CPS requirements and construct requirements modeling. To achieve this goal, we focus on benchmark construction, empirical evaluation, impact of the number of shots, and hallucinations type analysis. We believe these questions have great potential to provide insights and value for subsequent CPS requirements analysis and modeling by LLMs.

\textbf{Internal Validity} concerns the threats to how we perform our study. The first threat is related to the benchmark construction (Section \ref{sec:benchmark}). To construct the benchmark, we manually annotated the requirements texts. We acknowledge that these annotations are somewhat subjective. To mitigate this threat, we provided annotators with an annotation guide and held three meetings to learn about problem diagrams and the annotation tool. Then each annotator independently annotated the benchmark and each label was cross-validated. Besides, we also calculated inter-annotator consistency scores. The second threat relates to the setup of LLMs when addressing RQ1 and RQ2. LLMs may show different performances under different decode strategies or inference frameworks. To mitigate this threat, we set LLMs as greedy decoding and the same inference framework. The third threat concerns the manual review of ground truth and predictions when addressing RQ3. It is hard to ensure that one person's review results are complete. To mitigate this threat, we construct a team to review them.

\textbf{External Validity} concerns the threats to generalize our findings. 
The first threat is the representativeness of our benchmark. To mitigate this threat, the requirements documents in our benchmark include various types of CPSs, covering a wide range of application domains (\eg transportation, military, telecommunications, and astronomy). This ensures that the collected requirements documents can represent the diversity of CPSs requirements modeling. The second threat is the selection of LLMs. We select the latest version of the LLMs with around 7 billion parameters released by well-known companies or organizations. It is to ensure the practicality and efficiency of the LLMs in real-world applications as LLMs with excessive parameters might not be feasible for organizations to deal with requirements tasks due to the limitation of computing resources. 



\subsection{Future Directions} 
Current LLMs have indeed shown potential in requirements modeling for CPSs. We believe that future work can be conducted in-depth from the following aspect. (1) \textbf{Develop CPS-specific LLMs}: customized and refined LLMs for CPSs requirement analysis and understanding, incorporating extensive CPSs domain knowledge during the pre-training phase to enhance LLM’s understanding of CPSs requirements documents. (2) \textbf{Enhance LLMs with modeling knowledge}: address the current shortfall of LLMs in identifying specialized concepts in requirement models by devising retrieval enhancement strategies or instruction fine-tuning datasets to inject requirement modeling knowledge. (3) \textbf{Integrate knowledge of multiple LLMs}: explore methods to integrate advantage of different LLMs to construct a mixture of experts for requirements modeling.

\section{Related Works} \label{sec:relatedworks}

\textbf{CPSs requirements modeling:} Model-Driven Software Development (MDSD)~\cite{volter2013model} has become a leading paradigm for developing CPSs and verifying requirements~\cite {geismann2020systematic}. Jin~\cite{jin_environment_2018} introduced an environment-based modeling approach which is highly potential to capture and express the requirements of CPSs. Helal et al.~\cite{helal2024towards} proposed a formal requirements modeling approach for CPSs. These works rely on human understanding to identify modeling elements and construct requirements modeling. 

Several approaches have been explored to convert natural language requirements to specific models such as class model~\cite{deeptimahanti2009automated}, feature model~\cite{davril2013feature}, and use case model~\cite{elbendak2011parsed}. These works can be divided into four types. (1). Approaches based on rules~\cite{yue2015atoucan}~\cite{nguyen2015rule} designed heuristic rules to construct requirements models from requirements in natural language, but these rules designed by these works are difficult to transfer. (2) Approaches based on text processing~\cite{das2024extracting}~\cite{elbendak2011parsed} use syntactic and semantic analysis of the requirement text to identify modeling elements. They are suitable for identifying explicit basic requirements, but they are difficult to analyze the interactions between them~\cite{wang2021survey}. (3) Approaches based on deep learning~\cite{sainani2020extracting}~\cite{zhou2022assisting} mainly concentrate on extracting
requirements from software contracts or software reviews and pay less attention to suit the requirements modeling. (4) Approaches based on LLMs~\cite{camara2023assessment}~\cite{chen2023automated} rely on the understanding capabilities of LLMs to extract modeling elements from requirements text. However, these works based on LLMs~\cite{camara2023assessment}~\cite{chen2023automated} only focus on a single concern (\eg use case or sequence). Unlike them, this paper focuses on CPS requirements and aims to construct problem diagrams from CPS requirements documents, which requires understanding multiple concerns (\eg environment and interactions).

\textbf{Evaluation of LLMs on requirements modeling:} LLMs have demonstrated excellent performance in natural language understanding. Researchers have begun to evaluate the capability of LLMs to understand requirements and build requirements models. Cámara et al.~\cite{camara2023assessment} investigated the capabilities of ChatGPT to build UML models through 6 cases. Ferrari et al.~\cite{ferrari2024model} measured the performance of ChatGPT to generate sequence models through multiple cases. Chen et al.~\cite{chen2023automated} evaluated the potential of GPT4 for generating goal models from NL descriptions of the problem context. Ruan et al.~\cite{RuanCJ23} evaluated the ability of ChatGPT to model CPS requirements through a digital home system case. However, the evaluated cases and LLMs in these works~\cite{camara2023assessment}~\cite{ferrari2024model}~\cite{chen2023automated}~\cite{RuanCJ23} are limited, making it difficult to comprehensively assess the capabilities of LLMs. Compared with them, this paper conducts a systematic evaluation of advanced LLMs through multiple CPS systems.

\section{Conclusion} \label{sec:conclusion}
This paper presents an empirical evaluation of the CPS requirements modeling abilities of LLMs. This work formulates the requirements modeling into two tasks and constructs a benchmark for requirements modeling of CPSs named {\sc CPSBench}. Using the benchmark, an extensive evaluation of advanced LLMs is conducted, gaining some insights into their strength and limitations. To further enhance the ability of LLMs for future research, we establish a taxonomy of LLMs hallucinations in requirements modeling and discuss future directions.


\useunder{\uline}{\ul}{}

\normalem
\balance
\bibliographystyle{IEEEtran}
\bibliography{myrefs}

\begin{thebibliography}{10}
\providecommand{\url}[1]{#1}
\csname url@samestyle\endcsname
\providecommand{\newblock}{\relax}
\providecommand{\bibinfo}[2]{#2}
\providecommand{\BIBentrySTDinterwordspacing}{\spaceskip=0pt\relax}
\providecommand{\BIBentryALTinterwordstretchfactor}{4}
\providecommand{\BIBentryALTinterwordspacing}{\spaceskip=\fontdimen2\font plus
\BIBentryALTinterwordstretchfactor\fontdimen3\font minus \fontdimen4\font\relax}
\providecommand{\BIBforeignlanguage}[2]{{%
\expandafter\ifx\csname l@#1\endcsname\relax
\typeout{** WARNING: IEEEtran.bst: No hyphenation pattern has been}%
\typeout{** loaded for the language `#1'. Using the pattern for}%
\typeout{** the default language instead.}%
\else
\language=\csname l@#1\endcsname
\fi
#2}}
\providecommand{\BIBdecl}{\relax}
\BIBdecl

\bibitem{nguyen2017model}
P.~H. Nguyen, S.~Ali, and T.~Yue, ``Model-based security engineering for cyber-physical systems: A systematic mapping study,'' \emph{Information and Software Technology}, vol.~83, pp. 116--135, 2017.

\bibitem{menghi2020approximation}
C.~Menghi, S.~Nejati, L.~Briand, and Y.~I. Parache, ``Approximation-refinement testing of compute-intensive cyber-physical models: An approach based on system identification,'' in \emph{Proceedings of the ACM/IEEE 42nd International Conference on Software Engineering}, 2020, pp. 372--384.

\bibitem{xu2023semantics}
X.~Xu, J.-P. Talpin, S.~Wang, B.~Zhan, and N.~Zhan, ``Semantics foundation for cyber-physical systems using higher-order utp,'' \emph{ACM Transactions on Software Engineering and Methodology}, vol.~32, no.~1, pp. 1--48, 2023.

\bibitem{mandrioli2023stress}
C.~Mandrioli, S.~Y. Shin, M.~Maggio, D.~Bianculli, and L.~Briand, ``Stress testing control loops in cyber-physical systems,'' \emph{ACM Transactions on Software Engineering and Methodology}, vol.~33, no.~2, pp. 1--58, 2023.

\bibitem{frepa}
J.~Feng, W.~Miao, H.~Zheng, Y.~Huang, J.~Li, Z.~Wang, T.~Su, B.~Gu, G.~Pu, M.~Yang, and J.~He, ``{FREPA:} an automated and formal approach to requirement modeling and analysis in aircraft control domain,'' in \emph{28th {ACM} Joint European Software Engineering Conference and Symposium on the Foundations of Software Engineering}, 2020, pp. 1376--1386.

\bibitem{stadler2022flexible}
M.~Stadler, M.~Vierhauser, A.~Garmendia, M.~Wimmer, and J.~Cleland-Huang, ``Flexible model-driven runtime monitoring support for cyber-physical systems,'' in \emph{Proceedings of the ACM/IEEE 44th International Conference on Software Engineering: Companion Proceedings}, 2022, pp. 350--351.

\bibitem{BouskelaFGJOTT22}
D.~Bouskela, A.~Falcone, A.~Garro, A.~Jardin, M.~Otter, N.~Thuy, and A.~Tundis, ``Formal requirements modeling for cyber-physical systems engineering: an integrated solution based on {FORM-L} and modelica,'' \emph{Requir. Eng.}, vol.~27, no.~1, pp. 1--30, 2022.

\bibitem{jackson2005problem}
M.~Jackson, ``Problem frames and software engineering,'' \emph{Information and Software Technology}, vol.~47, no.~14, pp. 903--912, 2005.

\bibitem{JinRE2019}
Z.~Jin, X.~Chen, Z.~Li, and Y.~Yu, ``{RE4CPS:} requirements engineering for cyber-physical systems,'' in \emph{27th {IEEE} International Requirements Engineering Conference, {RE}}.\hskip 1em plus 0.5em minus 0.4em\relax {IEEE}, 2019, pp. 496--497.

\bibitem{jin_environment_2018}
Z.~Jin, \emph{Environment {{Modeling}}-Based {{Requirements Engineering}} for {{Software Intensive Systems}}}.\hskip 1em plus 0.5em minus 0.4em\relax Massachusetts, USA: {Morgan Kaufmann}, 2018.

\bibitem{rajbhoj2023doctomodel}
A.~Rajbhoj, P.~Nistala, V.~Kulkarni, S.~Soni, and A.~Pathan, ``Doctomodel: automated authoring of models from diverse requirements specification documents,'' in \emph{2023 IEEE/ACM 45th International Conference on Software Engineering: Software Engineering in Practice}.\hskip 1em plus 0.5em minus 0.4em\relax IEEE, 2023, pp. 199--210.

\bibitem{zaki2022rcm}
A.~Zaki-Ismail, M.~Osama, M.~Abdelrazek, J.~Grundy, and A.~Ibrahim, ``Rcm-extractor: an automated nlp-based approach for extracting a semi formal representation model from natural language requirements,'' \emph{Automated Software Engineering}, vol.~29, no.~1, p.~10, 2022.

\bibitem{javed2021imer}
M.~Javed and Y.~Lin, ``imer: Iterative process of entity relationship and business process model extraction from the requirements,'' \emph{Information and Software Technology}, vol. 135, p. 106558, 2021.

\bibitem{gpt-3.5}
OpenAI, ``gpt-3.5-turbo,'' \url{https://platform.openai.com/docs/models/gpt-3-5}, 2023.

\bibitem{arora2024advancing}
C.~Arora, J.~Grundy, and M.~Abdelrazek, ``Advancing requirements engineering through generative ai: Assessing the role of llms,'' in \emph{Generative AI for Effective Software Development}.\hskip 1em plus 0.5em minus 0.4em\relax Springer, 2024, pp. 129--148.

\bibitem{luitel2024improving}
D.~Luitel, S.~Hassani, and M.~Sabetzadeh, ``Improving requirements completeness: Automated assistance through large language models,'' \emph{Requirements Engineering}, vol.~29, no.~1, pp. 73--95, 2024.

\bibitem{lutze2024generating}
R.~Lutze and K.~Waldh{\"o}r, ``Generating specifications from requirements documents for smart devices using large language models (llms),'' in \emph{International Conference on Human-Computer Interaction}.\hskip 1em plus 0.5em minus 0.4em\relax Springer, 2024, pp. 94--108.

\bibitem{fantechi2023inconsistency}
A.~Fantechi, S.~Gnesi, L.~Passaro, and L.~Semini, ``Inconsistency detection in natural language requirements using chatgpt: a preliminary evaluation,'' in \emph{2023 IEEE 31st International Requirements Engineering Conference}.\hskip 1em plus 0.5em minus 0.4em\relax IEEE, 2023, pp. 335--340.

\bibitem{camara2023assessment}
J.~C{\'a}mara, J.~Troya, L.~Burgue{\~n}o, and A.~Vallecillo, ``On the assessment of generative ai in modeling tasks: an experience report with chatgpt and uml,'' \emph{Software and Systems Modeling}, vol.~22, no.~3, pp. 781--793, 2023.

\bibitem{ferrari2024model}
A.~Ferrari, S.~Abualhaija, and C.~Arora, ``Model generation from requirements with llms: an exploratory study,'' \emph{arXiv preprint arXiv:2404.06371}, 2024.

\bibitem{chen2023automated}
K.~Chen, Y.~Yang, B.~Chen, J.~A.~H. L{\'o}pez, G.~Mussbacher, and D.~Varr{\'o}, ``Automated domain modeling with large language models: A comparative study,'' in \emph{2023 ACM/IEEE 26th International Conference on Model Driven Engineering Languages and Systems (MODELS)}.\hskip 1em plus 0.5em minus 0.4em\relax IEEE, 2023, pp. 162--172.

\bibitem{RuanCJ23}
K.~Ruan, X.~Chen, and Z.~Jin, ``Requirements modeling aided by chatgpt: An experience in embedded systems,'' in \emph{31st {IEEE} International Requirements Engineering Conference Workshops}, 2023, pp. 170--177.

\bibitem{al2018use}
A.~Al-Hroob, A.~T. Imam, and R.~Al-Heisa, ``The use of artificial neural networks for extracting actions and actors from requirements document,'' \emph{Information and Software Technology}, vol. 101, pp. 1--15, 2018.

\bibitem{nguyen2015rule}
T.~H. Nguyen, J.~Grundy, and M.~Almorsy, ``Rule-based extraction of goal-use case models from text,'' in \emph{Proceedings of the 2015 10th Joint Meeting on Foundations of Software Engineering}, 2015, pp. 591--601.

\bibitem{xie2021explanation}
S.~M. Xie, A.~Raghunathan, P.~Liang, and T.~Ma, ``An explanation of in-context learning as implicit bayesian inference,'' in \emph{The Twelfth International Conference on Learning Representations}, 2021.

\bibitem{web:code}
``Our code and dataset,'' \url{https://anonymous.4open.science/r/CPSBench-BED7}.

\bibitem{ferrari2017pure}
A.~Ferrari, G.~O. Spagnolo, and S.~Gnesi, ``Pure: A dataset of public requirements documents,'' in \emph{2017 IEEE 25th international requirements engineering conference}.\hskip 1em plus 0.5em minus 0.4em\relax IEEE, 2017, pp. 502--505.

\bibitem{mavridou2020ten}
A.~Mavridou, H.~Bourbouh, D.~Giannakopoulou, T.~Pressburger, M.~Hejase, P.-L. Garoche, and J.~Schumann, ``The ten lockheed martin cyber-physical challenges: formalized, analyzed, and explained,'' in \emph{2020 IEEE 28th International Requirements Engineering Conference}, 2020.

\bibitem{yang2022intelligent}
M.~Yang, B.~Gu, Z.~Duan, Z.~Jin, N.~Zhan, Y.~Dong, C.~Tian, G.~Li, X.~Dong, and X.~Li, ``Intelligent program synthesis framework and key scientific problems for embedded software,'' \emph{Chinese Space Science and Technology}, vol.~42, no.~4, pp. 1--7, 2022.

\bibitem{projection}
X.~Wang, X.~Chen, Z.~Jin, B.~Gu, and Y.~Qi, ``Pojection-based requirements analysis approach for embedded systems.'' \emph{Journal of Software}.

\bibitem{chatmodeler}
D.~Jin, Z.~Jin, X.~Chen, and C.~Wang, ``Chatmodeler: A human-machine collaborative and iterative requirements elicitation and modeling approach via large language models,'' \emph{Journal of Computer Research and Development}, vol.~61, no.~2, pp. 338--350, 2024.

\bibitem{vasiliev2020natural}
Y.~Vasiliev, \emph{Natural language processing with Python and spaCy: A practical introduction}.\hskip 1em plus 0.5em minus 0.4em\relax No Starch Press, 2020.

\bibitem{label}
Heartex, ``Label-studio,'' \url{https://labelstud.io/}, 2023.

\bibitem{artstein2017inter}
R.~Artstein, ``Inter-annotator agreement,'' \emph{Handbook of linguistic annotation}, pp. 297--313, 2017.

\bibitem{mchugh2012interrater}
M.~L. McHugh, ``Interrater reliability: the kappa statistic,'' \emph{Biochemia medica}, vol.~22, no.~3, pp. 276--282, 2012.

\bibitem{abs-2404-00152}
M.~Munnangi, S.~Feldman, B.~C. Wallace, S.~Amir, T.~Hope, and A.~Naik, ``On-the-fly definition augmentation of llms for biomedical {NER},'' \emph{CoRR}, vol. abs/2404.00152, 2024.

\bibitem{wang2023gpt}
S.~Wang, X.~Sun, X.~Li, R.~Ouyang, F.~Wu, T.~Zhang, J.~Li, and G.~Wang, ``Gpt-ner: Named entity recognition via large language models,'' \emph{arXiv preprint arXiv:2304.10428}, 2023.

\bibitem{wu2019active}
D.~Wu, C.-T. Lin, and J.~Huang, ``Active learning for regression using greedy sampling,'' \emph{Information Sciences}, vol. 474, pp. 90--105, 2019.

\bibitem{openai2023gpt}
R.~OpenAI \emph{et~al.}, ``Gpt-4 technical report,'' \emph{ArXiv}, vol. 2303, p. 08774, 2023.

\bibitem{Qwen}
J.~Bai, S.~Bai, Y.~Chu, Z.~Cui, K.~Dang, X.~Deng, Y.~Fan, W.~Ge, Y.~Han, F.~Huang, B.~Hui, L.~Ji, M.~Li, J.~Lin, R.~Lin, D.~Liu, G.~Liu, C.~Lu, K.~Lu, J.~Ma, R.~Men, X.~Ren, X.~Ren, C.~Tan, S.~Tan, J.~Tu, P.~Wang, S.~Wang, W.~Wang, S.~Wu, B.~Xu, J.~Xu, A.~Yang, H.~Yang, J.~Yang, S.~Yang, Y.~Yao, B.~Yu, H.~Yuan, Z.~Yuan, J.~Zhang, X.~Zhang, Y.~Zhang, Z.~Zhang, C.~Zhou, J.~Zhou, X.~Zhou, and T.~Zhu, ``Qwen technical report,'' \emph{CoRR}, vol. abs/2309.16609, 2023.

\bibitem{LLama3}
OpenAI, ``Llama3-7b,'' \url{https://ai.meta.com/}, 2023.

\bibitem{gemma}
{Gemma}Team, ``Gemma: Open models based on gemini research and technology,'' \emph{arXiv:2403.08295}, 2024.

\bibitem{glm}
A.~Zeng, X.~Liu, Z.~Du, Z.~Wang, H.~Lai, M.~Ding, Z.~Yang, Y.~Xu, W.~Zheng, X.~Xia \emph{et~al.}, ``Glm-130b: An open bilingual pre-trained model.''

\bibitem{Mistral-7B}
A.~Q. Jiang, A.~Sablayrolles, A.~Mensch, C.~Bamford, D.~S. Chaplot, D.~de~Las~Casas, F.~Bressand, G.~Lengyel, G.~Lample, L.~Saulnier, L.~R. Lavaud, M.~Lachaux, P.~Stock, T.~L. Scao, T.~Lavril, T.~Wang, T.~Lacroix, and W.~E. Sayed, ``Mistral 7b,'' \emph{CoRR}, vol. abs/2310.06825, 2023.

\bibitem{shen2023promptner}
Y.~Shen, Z.~Tan, S.~Wu, W.~Zhang, R.~Zhang, Y.~Xi, W.~Lu, and Y.~Zhuang, ``{P}rompt{NER}: Prompt locating and typing for named entity recognition,'' in \emph{Proceedings of the 61st Annual Meeting of the Association for Computational Linguistics}, 2023, pp. 12\,492--12\,507.

\bibitem{wang2022matchprompt}
J.~Wang, L.~Zhang, J.~Liu, X.~Liang, Y.~Zhong, and Y.~Wu, in \emph{Proceedings of the 2022 Conference on Empirical Methods in Natural Language Processing}, 2022, pp. 7875--7888.

\bibitem{openaiapi}
OpenAI, ``Api access,'' \url{https://openai.com/index/openai-api/}, 2023.

\bibitem{chickering2002optimal}
D.~M. Chickering, ``Optimal structure identification with greedy search,'' \emph{Journal of machine learning research}, vol.~3, no. Nov, pp. 507--554, 2002.

\bibitem{web:vllm}
Berkeley, ``Vllm,'' \url{https://github.com/vllm-project/vllm}, 2024.

\bibitem{gao2021simcse}
T.~Gao, X.~Yao, and D.~Chen, ``{S}im{CSE}: Simple contrastive learning of sentence embeddings,'' in \emph{Proceedings of the 2021 Conference on Empirical Methods in Natural Language Processing}, 2021, pp. {6894--6910}.

\bibitem{douze2024faiss}
M.~Douze, A.~Guzhva, C.~Deng, J.~Johnson, G.~Szilvasy, P.-E. Mazar{\'e}, M.~Lomeli, L.~Hosseini, and H.~J{\'e}gou, ``The faiss library,'' \emph{arXiv preprint arXiv:2401.08281}, 2024.

\bibitem{gradio}
Gradio, ``Gradio,'' \url{https://www.gradio.app/}, 2023.

\bibitem{volter2013model}
M.~V{\"o}lter, T.~Stahl, J.~Bettin, A.~Haase, and S.~Helsen, \emph{Model-driven software development: technology, engineering, management}.\hskip 1em plus 0.5em minus 0.4em\relax John Wiley \& Sons, 2013.

\bibitem{geismann2020systematic}
J.~Geismann and E.~Bodden, ``A systematic literature review of model-driven security engineering for cyber--physical systems,'' \emph{Journal of Systems and Software}, vol. 169, p. 110697, 2020.

\bibitem{helal2024towards}
R.~Helal, A.~Seghiri, F.~Belala, and N.~Hameurlain, ``Towards a formal modeling approach for cyber-physical systems requirements,'' in \emph{Proceedings of the 2024 13th International Conference on Software and Computer Applications}, 2024, pp. 298--309.

\bibitem{deeptimahanti2009automated}
D.~K. Deeptimahanti and M.~A. Babar, ``An automated tool for generating uml models from natural language requirements,'' in \emph{2009 IEEE/ACM International Conference on Automated Software Engineering}.\hskip 1em plus 0.5em minus 0.4em\relax IEEE, 2009, pp. 680--682.

\bibitem{davril2013feature}
J.-M. Davril, E.~Delfosse, N.~Hariri, M.~Acher, J.~Cleland-Huang, and P.~Heymans, ``Feature model extraction from large collections of informal product descriptions,'' in \emph{proceedings of the 2013 9th joint meeting on foundations of software engineering}, 2013, pp. 290--300.

\bibitem{elbendak2011parsed}
M.~Elbendak, P.~Vickers, and N.~Rossiter, ``Parsed use case descriptions as a basis for object-oriented class model generation,'' \emph{Journal of Systems and Software}, vol.~84, no.~7, pp. 1209--1223, 2011.

\bibitem{yue2015atoucan}
T.~Yue, L.~C. Briand, and Y.~Labiche, ``atoucan: an automated framework to derive uml analysis models from use case models,'' \emph{ACM Transactions on Software Engineering and Methodology}, vol.~24, no.~3, pp. 1--52, 2015.

\bibitem{das2024extracting}
S.~Das, N.~Deb, A.~Cortesi, and N.~Chaki, ``Extracting goal models from natural language requirement specifications,'' \emph{Journal of Systems and Software}, vol. 211, p. 111981, 2024.

\bibitem{wang2021survey}
Y.~Wang and B.~J. Junwu~Chen, Xin~Xia, ``Intelligent requirements elicitation and modeling: A literature review,'' \emph{Journal of Computer Research and Development}, vol.~58, no.~4, pp. 683--705, 2021.

\bibitem{sainani2020extracting}
A.~Sainani, P.~R. Anish, V.~Joshi, and S.~Ghaisas, ``Extracting and classifying requirements from software engineering contracts,'' in \emph{2020 IEEE 28th international requirements engineering conference}, 2020, pp. 147--157.

\bibitem{zhou2022assisting}
Q.~Zhou, T.~Li, and Y.~Wang, ``Assisting in requirements goal modeling: a hybrid approach based on machine learning and logical reasoning,'' in \emph{Proceedings of the 25th International Conference on Model Driven Engineering Languages and Systems}, 2022, pp. 199--209.

\end{thebibliography}

\end{document}